\documentclass[preprint2]{aastex}
\usepackage{spr-astr-addons}
\usepackage{url}
\usepackage{subfig}
\let\captionbox=\undefined
\usepackage{caption} 
\RequirePackage{color}

\newcommand{\emaila}{tasfaw@einstein.physics.drexel.edu}

\begin{document}

\title{Modeling coronal magnetic field using spherical geometry: cases with several active regions}
\slugcomment{Modeling coronal magnetic field}
\shorttitle{Modeling coronal magnetic field}
\shortauthors{T.~Tadesse et al.}

\author{Tilaye Tadesse\altaffilmark{1,2}}
\affil{Department of Physics, Drexel University, Philadelphia, PA 19104-2875, U.S.A.}
\email{\emaila}
 \altaffiltext{2}{Addis Ababa University, College of Natural Sciences, Institute of Geophysics, Space Science, and Astronomy,
                         Po.Box 1176, Addis Ababa, Ethiopia}
\and 
\author{T. Wiegelmann\altaffilmark{3}}
\affil{Max Planck Institut f\"{u}r Sonnensystemforschung, Max-Planck Str. 2, D--37191 Katlenburg-Lindau, Germany}

\and
\author{K. Olson\altaffilmark{1}}
\affil{Department of Physics, Drexel University, Philadelphia, PA 19104-2875, U.S.A.}
\and
\author{ P. J. MacNeice\altaffilmark{4}}
\affil{NASA, Goddard Space Flight Center, Code 674, Greenbelt, MD 20771, U.S.A.}

\begin{abstract}
The magnetic fields in the solar atmosphere structure the plasma, store free magnetic energy and produce a wide variety 
of active solar phenomena, like flare and coronal mass ejections(CMEs). The distribution and strength of magnetic fields 
are routinely measured in the solar surface( photosphere). Therefore, there is considerable interest in accurately modeling 
the 3D structure of the coronal magnetic field using photospheric vector magnetograms. Knowledge of the 3D structure of 
magnetic field lines also help us to interpret other coronal observations, e.g., EUV images of the radiating coronal plasma. 
Nonlinear force-free field (NLFFF) models are thought to be viable tools for those task. Usually those models use Cartesian 
geometry. However, the spherical nature of the solar surface cannot be neglected when the field of view is large. In this work, 
we model the coronal magnetic field above multiple active regions using NLFFF extrapolation code using vector magnetograph 
data from the Synoptic Optical Long-term Investigations of the Sun survey (SOLIS)/ Vector Spectromagnetograph (VSM) as a boundary 
conditions. We compare projections of the resulting magnetic field lines solutions with their respective coronal EUV-images from 
the Atmospheric Imaging Assembly (SDO/AIA) observed on October 11, 2011 and November 13, 2012. This study has found that the NLFFF 
model in spherical geometry reconstructs the magnetic configurations for several active regions which agrees with observations. During 
October 11, 2011 observation, there are substantial number of trans-equatorial loops carrying electric current. 

\end{abstract}


\section{Introduction}
The geometry and dynamics of the solar corona are determined by the evolving magnetic field at the Sun’s 
surface (photosphere). Thus, magnetic fields are believed to play a dominant role for active phenomena 
carried out in the solar corona. In order to study solar eruptive phenomena, we have to understand how 
magnetic energy is stored in the pre-eruptive corona. Hence, the three dimensional (3D) structure of magnetic 
fields and electric currents in the pre-eruptive corona and the amount of free energy stored in the field 
have to determined \citep{schrijver:etal2005,Jiang:2012}. Unfortunately, reliable magnetic field measurements 
are still restricted to the level of the photosphere, where the inverse Zeeman effect in Fraunhofer lines is 
observable. Even if the measurement of magnetic fields in the solar corona has considerably improved in recent 
decades \citep{Lin:2000,Liu:2008}, further developments are needed before accurate data are routinely available. 

As an alternative to measurements in the solar atmosphere, magnetic field extrapolation have been implemented 
to calculate the field from the measured photospheric field using assumptions that the fields are force-free. 
In this model assumption, the corona magnetic forces are dominant that all non-magnetic forces like plasma 
pressure gradient and gravity can be neglected in the lowest order \citep{Gary}. This implies that, if
there are currents, they must be aligned with the magnetic field. To describe the equilibrium structure of the 
static coronal magnetic field, the force-free assumption is appropriate:
\begin{equation}
   (\nabla \times\textbf{B})=\alpha\textbf{B} \label{one}
\end{equation}
\begin{equation}
    \nabla \cdot\textbf{B}=0 \label{two}
 \end{equation}
 subject to the boundary condition
\begin{equation}
    \textbf{B}=\textbf{B}_{\textrm{obs}} \quad \mbox{on photosphere} \label{three}
 \end{equation}
where $\textbf{B}$ is the magnetic field and $\textbf{B}_{\textrm{obs}}$ is measured vector field on 
the photosphere. Equation~(\ref{one}) states that the currents are co-aligned with magnetic fields, where 
$\alpha$ is the torsion function that represents the proportionality between the electric current
density and the magnetic field. Equation~(\ref{two}) describes the absence of magnetic monopoles. 

Parameter $\alpha$ of Equation~(\ref{one}) can be a function of position, but the combination of Equation~(\ref{one}) 
and Equation~(\ref{two}) ($\textbf{B}\cdot\nabla\alpha=0$) requires that $\alpha$ be constant along a given field line. 
$\alpha=0$ yields the current-free potential field. If $\alpha$ is constant everywhere in the volume under consideration, 
the field is called linear force-free field (LFFF), otherwise it is nonlinear force-free field (NLFFF) 
\citep{Wiegelmann04,Wheatland:2009,Wheatland:2011,Wiegelmann:2012,Wiegelmann:2012W,Malanushenko:2012}. 
Potential and linear force-free fields can be used as a first step to model the general structure of magnetic 
fields in the solar corona, the use of nonlinear force free fields is essential to understand eruptive 
phenomena. Practically there are both observational and theoretical reasons which suggest that the pre-eruptive magnetic 
fields are non-linear force-free fields. For details of those models we direct the readers to \citet{Wiegelmann:2012W}.

Most of the NLFFF modeling codes are implemented in the Cartesian geometry. Therefore those codes are not well suited 
for larger domains, since the spherical nature of the solar surface cannot be neglected when the field of view is 
large. \citet{DeRosa} has studied that different NLFFF models have different field line configurations in the coronal 
volume with widely varying estimates of the magnetic free energy. The study suggested that the main reasons for that 
problem are (1) non-force free nature of the field within the photosphere, (2) the uncertainties on vector-field measurements, 
particularly on the transverse component, and (3) the needs of large domain that can capture the connections of an active 
region to its surroundings to accommodate more electric currents associated with the full active regions of interest. Solar 
Dynamics Observatory (SDO) mission has made repeated observations of large, almost global scale events in which large 
scale connection between active regions may play fundamental role. Therefore, this needs motivate us to implement a NLFFF 
procedure in spherical geometry to large scale boundary data from the Helioseismic and Magnetic Imager (HMI) on board 
SDO and Synoptic Optical Long-term Investigations of the Sun survey (SOLIS)/  Vector Spectromagnetograph (VSM) 
\citep{Wiegelmann07,tilaye09,Tilaye:2010,Tilaye:2012,Tilaye:2012a,Guo:2012,Tadesse:2013}

Coronal magnetic fields manifest themselves in X-rays and EUV images in the shapes of coronal loops. Those loops provide 
an additional constraint which is not currently being used due to the mathematical complications of incorporating such input 
into numerical models \citep{Malanushenko:2012}. The 3D structure of magnetic loops from NLFFF models 
help us for modeling of plasma loops, and understanding coronal heating and plasma flows along the loops. In this work, 
we apply our spherical NLFFF procedure to a group of active regions observed on October 11, 2011 and November 
13, 2012 using SOLIS/VSM data. During the former observation, there were six active regions (ARs 11610, 11611, 
11612, 11613, 11614 and 11615) and there were three vertically aligned active regions in later one. In our previous 
work \citet{Tilaye:2012a}, we have studied the connectivity between three neighboring active regions which were 
horizontally aligned. Here, we compare the extrapolated magnetic loops with their respective extreme ultraviolet 
(EUV) observations by the Atmospheric Imaging Assembly (AIA) on board SDO.
\section{Method}
This study requires extrapolating the three-dimensional NLFFF coronal fields from the photospheric boundary data. The
photospheric vector magnetograms, obtained by the Synoptic Optical Long-term Investigations of the Sun survey (SOLIS)/  
Vector Spectromagnetograph (VSM) are used as the boundary conditions. Meanwhile, those measured data are inconsistent 
with the above force-free assumption. Therefore, one has to apply some transformations to these data before nonlinear 
force-free extrapolation codes can be applied. This procedure is known as preprocessing. This preprocessing scheme 
removes forces and torques from the boundary and approximates the photospheric magnetic field to the low plasma-$\beta$
force-free chromosphere. For detailed descriptions of the method we direct the readers to \citet{Wiegelmann06sak} for 
Cartesian geometry and \citet{tilaye09} for spherical one.

We solve the force-free equations (\ref{one})(with the form $\nabla \times\textbf{B})\times\textbf{B}=0$) and 
(\ref{two}) using optimization principle \citep{Wheatland00,Wiegelmann04} in spherical geometry 
\citep{Wiegelmann07,tilaye09,Tadesse:2013,Tadesse:2013a}. The method minimizes a joint measure ($L_\mathrm{\omega}$) 
of the normalized Lorentz forces and the divergence of the field throughout the volume of interest, $V$. 
Optimization procedure is spherical geometry is given by:
\begin{equation}L_{\omega}=L_{f}+L_{d}+\nu L_{photo} \label{4}
\end{equation}
\begin{displaymath} L_{f}=\int_{V}\omega_{f}(r,\theta,\phi)B^{-2}\big|(\nabla\times {\textbf{B}})\times
{\textbf{B}}\big|^2  r^2\sin\theta dr d\theta d\phi
\end{displaymath}
\begin{displaymath}L_{d}=\int_{V}\omega_{d}(r,\theta,\phi)\big|\nabla\cdot {\textbf{B}}\big|^2
  r^2\sin\theta dr d\theta d\phi
\end{displaymath}
\begin{displaymath}L_{photo}=\int_{S}\textbf{B}_{d}\cdot\textbf{W}(\theta,\phi)\cdot\textbf{B}_{d} \,r^{2}\sin\theta d\theta d\phi
\end{displaymath}
where
$ \textbf{B}_{d}=\big(\textbf{B}-\textbf{B}_{obs}\big)$. $L_{f}$ and $L_{d}$ measure how well the force-free Eqs.~(\ref{one}) 
and divergence-free (\ref{two}) conditions are fulfilled, 
respectively, and both $\omega_{f}(r,\theta,\phi)$ and $\omega_{d}(r,\theta,\phi)$ are weighting functions. For detailed description 
of the use of weighting functions, we direct the readers to \citet{Wiegelmann04}. The term $L_{photo}$ is surface integral over the 
photosphere which allows us to relax the field on the photosphere towards force-free solution without to much deviation from the original 
surface field data, $\textbf{B}_{obs}$. In this integral, $\textbf{W}(\theta,\phi)$ is a space-dependent diagonal matrix which
gives different weights for observed surface field components depending on its relative accuracy in measurement. 

The full inversion of SOLIS data is performed in the framework of Milne-Eddington model (ME)\citep{Unno:1956}. The inversion is made 
only for pixels whose polarization signal is above a selected threshold. Pixels with polarization below threshold are left undetermined. 
These data gaps represent a major difficulty for existing magnetic field extrapolation schemes. Therefore, those missing data is considered 
most inaccurate and are taken account of by setting $\textbf{W}(\theta,\phi)$ to zero in all elements of the matrix. For the detailed 
description of the method we direct the readers to \citet{Wiegelmann10} and \citet{Tilaye:2010} for Cartesian and spherical geometry, 
respectively.

\section{Results}
\begin{figure*}[htp]
   \centering
\includegraphics[viewport=10 30 850 815,clip,height=17.4cm,width=18.5cm]{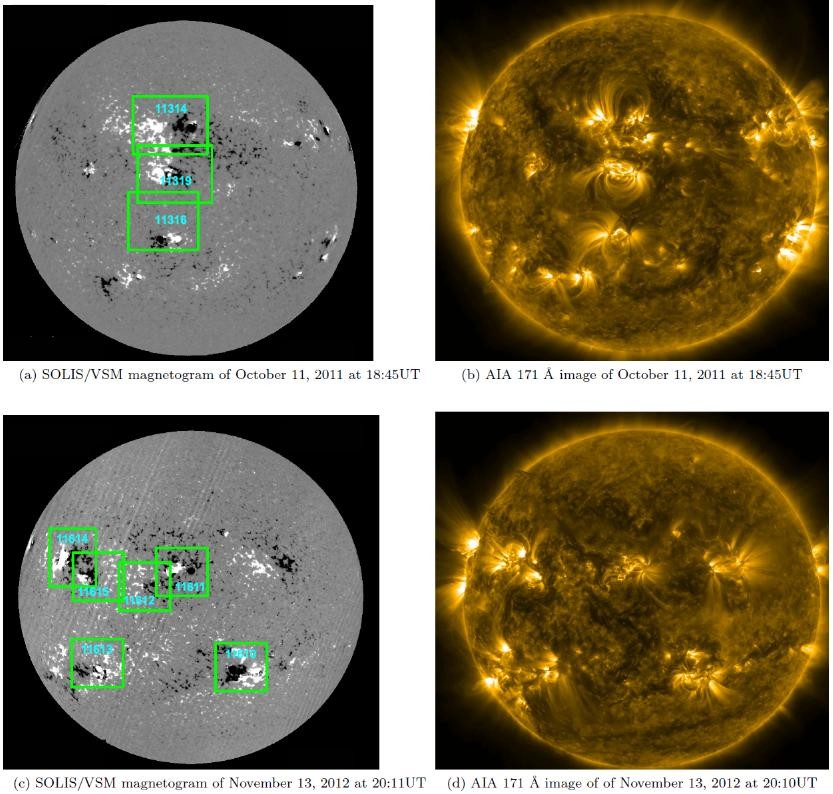}
 
\caption{a) Full-disk SOLIS/VSM magnetogram of October 11, 2011 at 18:45UT, b) Full-disk AIA 171 \AA{} image of October 11, 2011 at 18:45UT, c) 
Full-disk SOLIS/VSM magnetogram of November 13, 2012 at 20:11UT, d) Full-disk AIA 171 \AA{} image of of November 13, 2012 at 20:10UT. The blue 
boxes show the active regions observed.}\label{fig1}
 \end{figure*}

In this study, we have used our spherical NLFFF optimization procedure to a group of active regions observed  by SOLIS on October 11, 
2011 and November 13, 2012. During the former observation, there were six active regions (ARs 11610, 11611, 11612, 11613, 11614 and 11615) and 
there were three vertically aligned active regions(ARs 11314, 11316, and 11319) in later one (see Fig. \ref{fig1}a and b). In order 
to accommodate the connectivity between those ARs and their surroundings, we adopt a non uniform spherical grid $r$, $\theta$, and $\phi$ 
in the direction of radius, latitude, and longitude, respectively. Before applying our spherical NLFFF, we preprocessed the two 
photospheric vector field data sets using our spherical preprocessing routine. 

There are no vector magnetic field measurements for the side and top boundaries of a localized domain. Therefore, we have to make 
assumptions about these fields before performing a NLFFF extrapolation. We assumed the lateral and upper boundaries of the 
computational domain as current-free. In order to initialize our NLFFF code, we calculated potential field from those data sets 
using their respective preprocessed radial field components ($\textbf{B}_{r}$) using spherical harmonic expansion . The 
computational box is a wedge-shaped volume $V$ with six boundary surfaces (four lateral side boundaries, top and 
photospheric boundaries). This box enables us to study the connectivities between active regions and their surroundings 
for the large field-of-views.
\begin{figure*}[htp]
   \centering
\includegraphics[bb=10 10 850 815,clip,height=21.6cm,width=23.8cm]{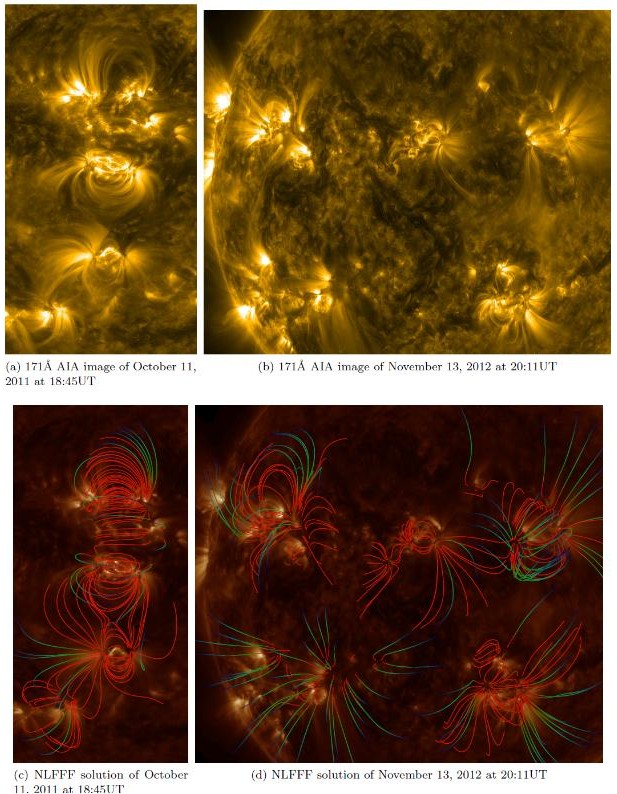}

\caption{a) SDO/AIA 171 \AA{} image on October 11, 2011 at 18:45UT, b) SDO/AIA 171 \AA{} image on  November 13, 2012 at 20:10UT, c) NLFFF model field 
lines of October 11, 2011 at 18:45UT overlaid on AIA 193\AA{} image and d) NLFFF model field lines of November 13, 2012 at 20:11UT overlaid on AIA 193\AA{} 
image. Green and red lines represent open and closed magnetic field lines.}\label{fig2}
 \end{figure*}

For computing the NLFFF field solutions for the two data sets, we use the preprocessed surface vector boundary fields with radial, 
longitudinal, and latitudinal components. We minimize the functional $L_{\omega}$ of Eq.~(\ref{4}). We implement the new term 
$L_{photo}$ in Eq.~(\ref{4}) to treat those data gaps in SOLIS \citep{Wiegelmann10,Tilaye:2010}. For those pixels, 
for which $\textbf{B}_{obs}$ was successfully inverted, we allow deviations between the model field $\textbf{B}$ and the input fields 
observed $\textbf{B}_{obs}$ surface field using Eq.~(\ref{4}), so that the model field can be iterated closer to a force-free solution 
even if the observations are inconsistent. We have used Langrangian multiplier of $\nu=0.001$ to control the speed with which the lower 
boundary is injected during the NLFFF extrapolation.  

In order to compare our reconstructions with observation, we plot the selected field lines of the NLFFF solutions for the two data sets and 
we overlay the field lines with corresponding AIA 193 \AA{} image (see Figure~\ref{fig2}c and d). From Figure~\ref{fig2}c and d, one can see 
that the field lines of NLFFF model solutions are reconstructed in such away that they fit with the observation. There are spatial correspondence 
between the overall shape of the NLFFF field lines and the EUV loops for both observational dates. Those qualitative comparisons between NLFFF model 
magnetic field lines and the observed EUV loops of AIA images indicate that the NLFFF model provides a more consistent field for large field-of-views. 
Even if the qualitative nature is not ideal, it remains the best option in the absence of a more reliable quantitative comparison.
\begin{center}
\begin{table}
\caption{The magnetic energy associated with extrapolated 3D NLFFF field configurations from SOLIS data sets of the two dates.}
\label{table3}
\begin{tabular}{ccc}
 \hline \hline
Date & $E_{\mbox{total}}(10^{33}{\mbox{erg}})$& $E_{\mbox{free}}(10^{33}{\mbox{erg}})$\\
\hline
October 11, 2011 &$5.63$&$0.12$\\
November 13, 2012 &$9.72$&$0.61$\\
\hline
\end{tabular}
\end{table}
\end{center}
Once we have determined the 3D structure of the coronal magnetic field using the sophisticated numerical modeling, we can use those field solutions 
to calculate different quantities in the solar corona. We estimate the free magnetic energy to be the difference between the extrapolated NLFFF and 
the potential field with the same normal boundary conditions in the photosphere\citet{Regnier,Thalmann}. We therefore estimate the upper limit to 
the free magnetic energy associated with coronal currents of the form
\begin{equation}
E_\mathrm{free}=\frac{1}{8\pi}\int_{V}\Big(B_{nlff}^{2}-B_{pot}^{2}\Big)r^{2}sin\theta dr d\theta d\phi, \label{ten}
\end{equation}
Our result for the estimation of free-magnetic energy in Table~\ref{table3} shows that the NLFFF solutions have $2.13\%$ and 
$6.28\%$ more energy than the corresponding potential field solutions obtained from data observed on October 11, 2011 and November 13, 2012, 
respectively. 

\begin{figure}[htp!]
   \centering
\includegraphics[viewport=40 40 540 700,clip,height=10.5cm,width=7.5cm]{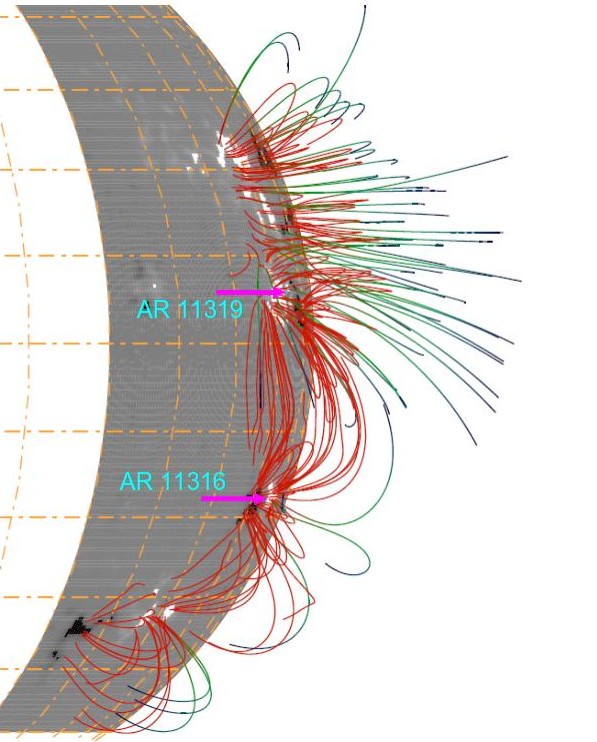}

\caption{Selected field lines of NLFFF model solutions of October 11, 2011 at 18:45UT rotated to limb.}\label{fig3}
 \end{figure}

In our previous work \citet{Tadesse:2013a}, we have studied the connectivity between two active regions one in the northern hemisphere 
and the other in the south. In this study we have used the same method to calculate the percentage of the total electric current shared 
between the AR 11319 in the northern hemisphere and AR 11316 in the southern hemisphere observed on October 11, 2011 at 18:45UT 
(see Fig.~\ref{fig3}). In order to quantify the percentage share in the electric current, we first identified those field lines 
carrying electric currents and emanating from AR 11319 and ending into AR 11316. The ratio of total electric current density flux 
due to those electric current carrying field lines connecting the two ARs to the total electric current density flux due to all 
field lines with current emanating from AR 11319 gives us the percentage share in the electric current between the two active 
regions. For this case study, we found that $23.31\%$ of positive/negative polarity of AR 11319 in the northern hemisphere is 
connected to positive/negative polarity of AR 11316 in southern hemisphere. There are more electric current connectivity between 
these ARs as they are closer to each other even if they are located in different hemispheres. Therefore, during this particular 
observation, there are substantial number of trans-equatorial loops carrying electric current. 
\section{Conclusion and outlook}
Coronal magnetic field modeling codes in Cartesian geometry are not well suited for larger domains, since the spherical 
nature of the solar surface cannot be neglected when the field of view is large. In this study, we have investigated the 
3D coronal magnetic field associated with two data sets observed on October 11, 2011 and  November 13, 2012 by analyzing 
SOLIS/VSM data using NLFFF model. During the former observation, there were six active regions (ARs 11610, 11611, 11612, 
11613, 11614 and 11615) and there were three vertically aligned active regions in later one. We have used our spherical NLFFF 
codes to compute the magnetic field solutions for large field of views with six active regions and three vertically aligned 
ones for the first time. 

For the comparison with observations, we overlayed the NLFFF model field solutions for the two data sets 
with their respective EUV loops from AIA. The study indicates that there are significant agreements between 
EUV loops and NLFFF model solutions. There are spatial correspondence between the overall shape of the NLFFF field lines and the 
EUV loops for both observational dates. Those qualitative comparisons between NLFFF model magnetic field lines and the observed EUV 
loops of AIA images indicate that the NLFFF model provides a more consistent field for large field-of-views. Even if the qualitative nature 
is not ideal, it remains the best option in the absence of a more reliable quantitative comparison.

Today, Solar Dynamics Observatory (SDO) mission has been repeated observations of large, almost global scale events in 
which large scale connection between active regions may play fundamental role. Therefore, it is useful to implement 
a NLFFF procedure in spherical geometry for use when large scale boundary data are available, such as from the 
Helioseismic and Magnetic Imager (HMI) on board SDO. In order to handle large resolution data from HMI, one can 
adapt the multi-grid technique to the spherical code to improve its performance.


\acknowledgments 
SOLIS/VSM vector magnetograms are produced cooperatively by NSF/NSO and NASA/LWS. 
The National Solar Observatory (NSO) is operated by the Association of Universities for Research in Astronomy, Inc., under cooperative agreement 
with the National Science Foundation. This work was supported by NASA grant NNX07AU64G and the work of T. Wiegelmann was supported by DLR-grant $50$ OC $453$  $0501$.

\end{document}